\documentclass[aps,pra,preprint,groupedaddress]{revtex4}

\newcommand{\be}{\begin{equation}}
\newcommand{\ee}{\end{equation}}

\begin{document}

\title{Bose-Einstein Condensation of Helium and Hydrogen inside Bundles of Carbon 
Nanotubes }

\author{F. Ancilotto$^1$,  M.M. Calbi$^2$, S.M. Gatica$^3$ and  M.W. Cole$^3$}

\affiliation{
$^1$INFM (UdR Padova and DEMOCRITOS National Simulation Center, Trieste, Italy)  and Dipartimento di Fisica ``G. Galilei'',
Universit\`a di Padova,\\ via Marzolo 8, I-35131 Padova, Italy.\\ 
$^2$ Department of Physics,  Southern Illinois University,
Carbondale, IL 62901-4401\\
$^3$  Department  of
Physics, Pennsylvania State
University, University Park, PA 16802}

\date{\today}

\begin{abstract}

Helium atoms or hydrogen molecules are believed to be strongly bound within the 
interstitial channels (between three carbon nanotubes) within a bundle of many 
nanotubes. The effects on adsorption of a nonuniform distribution of tubes are evaluated. 
The energy of a single particle state is the sum of a discrete transverse energy $E_t $
(that depends on the radii of neighboring tubes) and a quasicontinuous energy $E_z$ of relatively 
free motion parallel to the axis of the tubes. At low temperature, the particles occupy the 
lowest energy states, the focus of this study. The transverse energy attains a global minimum value ($E_t=E_{min}$) for 
radii near $R_{min}=9.95$ {\AA} for H$_2$ and 8.48 {\AA} for $^4$He. The density of states $N(E)$ near the lowest 
energy is found to vary linearly above this threshold value, i.e. $N(E)$ is proportional to $(E-
E_{min})$. As a result, there occurs a Bose-Einstein condensation of the molecules into the 
channel with the lowest transverse energy. The transition is characterized approximately 
as that of a four dimensional gas, neglecting the interactions between the adsorbed 
particles. The phenomenon is observable, in principle, from a singular heat capacity. The 
existence of this transition depends on the sample having a relatively broad distribution 
of radii values that include some near $R_{min}$.

\end{abstract}

\maketitle

\section{Introduction}

One of the most persistently interesting topics in condensed matter physics is Bose-
Einstein condensation (BEC). First postulated \cite{1} on the basis of an idealized model 
(noninteracting particles), BEC was subsequently argued by London to provide an 
explanation of superfluidity in $^4$He, which is a system consisting of relatively strongly 
interacting atoms \cite{2}. The London hypothesis has been  confirmed in both detailed 
calculations and careful neutron scattering experiments \cite{3,4}. More recently, BEC has 
been observed to occur in ultracold gases consisting of alkali or H atoms and in systems 
of excitons at low temperature $T$ \cite{5,6,7}. 

A logical candidate for the occurrence of BEC is para-H$_2$, which is the boson equilibrium 
form of hydrogen at low $T$. The apparent nonexistence of BEC for bulk H$_2$ is attributed to its 
crystallization below 14 K, which preempts BEC. Some three decades ago, Ginzburg and 
Sobyanin proposed a search for BEC and superfluidity of H$_2$ in confining geometries (for 
which crystallization  occurs at a lower temperature, if at all) \cite{8}. This suggestion stimulated a number of 
searches involving both theory and experiments \cite{9,10,11,12,13,14,15,16}. There has been at least one 
assertion of successful observation of BEC, involving the lack of damping of rotational motion of an impurity 
surrounded by hydrogen fluid within a small cluster \cite{9,14}.

In this paper, we describe a different geometry that is predicted to allow BEC to occur 
\cite{17,18}. This  host material is an ensemble of carbon nanotube bundles. A crucial role is played in this 
phenomenon by the presence of nonuniformity. In particular, we assume that there exists 
a fairly broad distribution of nanotube  radii present in the experimental sample. We focus our 
calculations on the spectrum of particles (either H$_2$ or $^4$He) confined within interstitial 
channels (ICs) formed by groups of three nanotubes. It has been argued that both of these 
species, if allowed access to such channels, are more strongly bound there than on the 
surface of graphite, the most strongly physisorbing planar surface \cite{19}. There is 
ambiguous experimental evidence concerning the relative binding energy of these species 
in the IC and in the grooves on the outside surface of the bundle \cite{20,21,22,23,24}. We interpret the 
very large isotope effect seen in isosteric heat data of Wilson et al \cite{22} to indicate particle  
localization within the ICs for their sample. The specific behavior is evidently sample-
dependent, just as our calculations are very sensitive to the assumptions that we make. 

Much recent research has been devoted to hydrogen within nanotubes bundles, partly due 
to the exciting prospects of technologies related to hydrogen storage and isotope 
separation \cite{25,26,27,28,29}. The vast majority of the theoretical research to date (including that of our 
group) has assumed a monodisperse distribution of tubes. The present study was 
stimulated by recent results of Shi and Johnson \cite{30}, who showed that the distribution of 
tube sizes present in actual samples leads to predictions that are quite different from those 
based on the monodisperse models, in better agreement with experimental data than the 
naive models' predictions. 

The outline of this paper is the following. Section II describes our calculations of the 
energy spectra of $^4$He and H$_2$ gases within ICs. Section III employs these results to 
compute a set of predictions related to BEC that are testable in principle. Section IV 
presents a further discussion of the model and draws conclusions.

\section{Energy spectra and density of states of adsorbed gases}

We focus here on those states having the lowest energy since they are the most important 
for the BEC phenomenon; higher energy  behavior  can be explored with the same 
approach but minor changes in the analysis are required. The energy of a particle 
confined within an IC is evaluated with the Schrodinger equation. Assuming that the 
particular IC is perfectly straight, this equation is separable; then the total energy is given 
by the relation:

\be
E_n(p)= E_n^{trans} + E^{long}(p)	
\ee
\label{1}

\be
E^{long}(p) = \frac{p^2}{2m}		
\ee
\label{2}
Here, the first term is the n-th eigenvalue of the Schrodinger equation describing motion 
transverse to the axis of the IC and the second term is the quasicontinuous kinetic energy 
of motion parallel to the axis. We assume for simplicity that the longitudinal kinetic 
energy is determined by the free particle mass; if the potential were corrugated the band 
mass should be used instead of m to characterize the low energy states \cite{31,32}. Such a 
substitution would lead to a straightforward change in the numerical and analytical results 
below.

In the present circumstance, the transverse eigenvalues are separated by energies (of 
order 100 K) that are much larger than relevant temperature scales \cite{33}. Hence, we need 
consider only the lowest eigenvalue $E_1^{trans}$ of transverse motion. We denote this quantity 
$E_t({\bf R})$, where the vector ${\bf R}=(R_1,R_2,R_3)$ has components equal to the radii of the tubes 
surrounding the IC. We refer to the domain of possible ICs as ${\bf R}$ space; a given IC is 
represented by one point in this space while a sample with many ICs is described by a 
cloud of points in this space. The single particle spectrum is now specified by the relation

\be
E({\bf R},p)= E_t({\bf R}) + \frac{p^2}{2m}	
\ee
\label{3}
Here, $E_t({\bf R})$  is computed following the procedure of Stan et.al.\cite{stan}. To get a sufficiently accurate dependence of $E_t$ on {\bf R}, one must include anisotropic and anharmonic contributions to the potential in the IC. We note that alternative parametrizations of the potential have been proposed. Table 1 presents one such alternative result for $E_t$. The qualitative behavior presented in the remainder of this paper is the same for the case of that alternative potential.  

Of considerable interest is the density of states $h(E,{\bf R})$ for a single IC. For a specific set 
of adjacent tubes (${\bf R}$), this function is given by a sum over all momenta, which may be 
replaced by an integral:

\be
h(E,{\bf R}) =\sum_p \delta[E- E({\bf R},p)] = \frac{L}{\pi\hbar}(\frac{m}{2})^{1/2} 
\frac{1}{(E- E_t({\bf R}))^{1/2}}
\, \Theta[E- E_t({\bf R})]	
\ee
\label{4}
Here $L$ is the length of the channel and the usual procedure for quantizing the 1D motion 
has been followed. The inverse square root dependence on the energy above threshold is 
characteristic of 1D motion. The last factor involves the Heaviside step function $\Theta(x)$, 
which is unity here for $E> E_t({\bf R})$ and zero otherwise.

Now let us consider an ensemble of nanotubes. The lowest energy states (those 
dominating the low temperature behavior) are concentrated near the minimum of the 
function $E_t({\bf R})$. As might be expected, the global minimum of $E_t({\bf R})$ is found for the 
symmetric case, $\{R_i\}=R_{min}$. We define the vector specifying this IC as ${\bf R}_{min}= R_{min} (1,1,1) $
and $E_{min}$ as the corresponding energy. The numerical results for these quantities appear in 
Table 1. 

The behavior of the function $E_t({\bf R})$ near this minimum is remarkable. If one considers 
only symmetric ICs (bounded by tubes of identical radii), the energy varies rapidly as a 
function of the difference $|{\bf R}- {\bf R}_{min} |$. However, the spectrum of the very lowest energy 
states is dominated by {\it asymmetric} ICs for which one of the radii equals $R_{min}$ and the two 
others differ from $R_{min}$ by equal, but opposite amounts, e.g. ${\bf R}= R_{min} (1,1+x,1-x)$, where 
$x\ll 1$. The energy varies extremely slowly with $x$ in such a case. This  is depicted for H$_2$ in 
Fig. 1, which shows the energy variation in a plane within ${\bf R}$ space that contains both this 
variable $x$ line and the diagonal $(1,1,1)$ line. (Analogous behavior occurs for the case of $^4$He). 
One observes in the figure a very narrow valley of low-lying states 
of variable $x$, along the $(0,1,-1)$ direction, where the origin is shifted to ${\bf R}_{min}$. Similar 
small gradient behavior of $E_t({\bf R})$ is found along equivalent permutations of this direction. 
Note that the higher energy contours are extended, nearly straight lines perpendicular to 
the diagonal. These latter contours imply an energy dependence of the density of states 
that is qualitatively different from that based on contours of lower energy (near the 
valleys).

We next evaluate the transverse density of states for a given experimental sample of 
tubes. We denote by $\nu({\bf R})$ the  density distribution of ICs, defined so that $\nu({\bf R}) d{\bf R}$ is the number 
of ICs present in the sample within a volume $d{\bf R}$ in ${\bf R}$ space. In this paper, we focus on low energy 
behavior corresponding to such close proximity to ${\bf R}_{min}$ that we may replace $\nu({\bf R})$ by 
$\nu ({\bf R}_{min})$ ; future studies will address more general situations for which that approximation is not adequate. In the following discussion, we assume that $\nu({\bf R})$ is a sufficiently broad 
distribution that ICs near this minimum exist in sufficient numbers to treat adsorption statistically. 
Moreover, we assume that the set of tube sizes forms a continuum. The former 
assumption is well justified in experimental samples produced to date involving large 
numbers of tubes, which typically have a dispersion in radius values of order 20\% 
\cite{34,35}. The latter (continuum) assumption is justified by the following argument. Near 
radius 1 nm, there are many values of the chiral indices that yield quite similar radii; the 
mean spacing between successive values of the radius  is quite small,  0.002 nm. In addition, the ICs 
experience small perturbations due to their environment; in particular, the ICs at the 
center of a bundle are expected to be compressed relative to those near the perimeter.
A model calculation (unpublished) yields an expression for the difference $\delta$ 
between the tube-tube separation in an infinite rope and that ($R_1$) of an isolated 
nanotube pair; apart from a constant of order one, the result is $\delta= [ C
\lambda^2]/[k R_1^6]$, where $C$ is the van der Waals-London intercarbon
interaction coefficient ($C\sim$ 20 eV-\AA$^6$), $\lambda$ is the 1D density of carbon atoms
in a nanotube ($\lambda\sim$15{\AA}$^{-1}$), $R_1\sim$17 {\AA} is the spacing between tube centers and $k$=10$^{-3}$ eV/\AA$^3$ is the force constant per unit length
associated with neighboring tubes' interactions  \cite{Mizel1998}. The result, $\delta\sim$0.15{\AA},   
represents a nearly 5\% compression of the lattice constant of ropium relative
to the separation of an isolated pair. This  shift in separation is consistent
with estimates of the analogous shift in breathing mode frequency \cite{?}.
 On 
the basis of these arguments, the assumption of a continuum of possible radii appears 
appropriate. 

The transverse density of states $g(E)$ for a given sample of tubes is expressed by the relation

\be
g(E) = \int d{\bf R}\, \nu({\bf R}) \,\delta[E - E_t({\bf R})]
\ee
\label{5}
The total density of states $N(E)$ is found by summing over all states of the ICs present in the 
sample:

\be
N(E)= \int d{\bf R} \sum_p \delta[E- E({\bf R},p)] \,\nu({\bf R})	
\ee
\label{6}

\be
N(E) = \frac{L}{\pi\hbar}(\frac{m}{2})^{1/2} \int_{E_{min}}^E dE_t\, g(E_t) \,[E- E_t]^{-1/2}	
\label{7}
\ee
Note that in the special case of a perfectly uniform distribution of $N_{IC}$ identical ICs, 
$\nu({\bf R})=N_{IC} \delta({\bf R}-{\bf R}_0)$ ,  yielding $g(E) = N_{IC}\delta[E - E_t({\bf R}_0)]$; then $N(E)$ is precisely $N_{IC}$ times 
the single tube's density of states, $h(E,{\bf R}_0)$, given in Eq. 4 above, as expected.	

Figs. 2 and 3 present the transverse densities of states $g(E_t)$ for He and H$_2$, respectively, 
computed numerically from a distribution of 600$^3$ tubes with radius values spread  uniformly over  the 
interval 8.5 {\AA} to 11.5 {\AA}. In both cases, the behavior of $g(E)$ at low energy is characterized by a 
power law dependence on energy above threshold : $g_{low}(E) \sim (E- E_{min})^{1/2}$. At higher 
energy, there is a decrease of $g(E)$  with energy, fit well by the expression $g_{high}(E) \sim (E- E_{min})^{-1/2}$. 
The initial behavior is explained by the following argument. At low energy, the density of 
states is obtained by counting the number of ICs whose energy $E_t({\bf R})$ lies in an interval 
between $E$ and $E+dE$. Assume for simplicity that the isoenergy surfaces are spherical 
surfaces centered about the point ${\bf R}= {\bf R}_{min}$ (i.e. $E-E_{min} =(k/2)| {\bf R}-{\bf R}_{min}|^2$). Then $g(E)$ is 
obtained from the number of points in the shell between spherical surfaces 
(corresponding to $E$ and $E+dE$) centered at ${\bf R}_{min}$. The result of this simple model is then

\be
g_{low}(E) = \nu({\bf R}_{min})\, 4\pi\, (\frac{2}{k^3})^{1/2}\, (E-E_{min})^{1/2}		
\label{8}
\ee
This dependence on the square root of the energy turns out to be valid even in the 
extreme anisotropic case of interest here \cite{36}. The result then is that $k^3$ is replaced by 
$k_{anis}^3=k_x k_t^2$, where $x$ denotes the (1,1,1) direction in ${\bf R}$ space and $t$ 
denotes the two directions 
transverse to that.

Note that this square root energy dependence of $g_{low}(E)$ coincides with that of the density 
of states of a 3D gas. The higher energy behavior ($1/\sqrt{E}$) of $g_{high}(E)$ is quite different. The relevant 
high energy regime in Fig. 1 is that where the energy surfaces in ${\bf R}$ space are 
perpendicular to the $(1,1,1)$ diagonal. Since this variation is essentially a 1D dependence, 
the resulting high energy form of $g_{high}(E)$ is that of a 1D system, as in Eq. 4 above; this 
gives rise to a behavior $g_{high}(E) \sim (E-E_{min})^{-1/2}$ (as is observed in Figs. 2 and 3).
We remark that this energy-dependent variation in effective dimensionality has analogs in electronic band structures, where wave vector replaces the ${\bf R}$ variable as the source of the unusual dependence. \cite{footnote}

Figs. 2 and 3, lower panels, depict the total densities of states $N(E)$ derived from Eq. \ref{7}. In 
view of the power law forms (e.g. Eq. \ref{8}) of $g(E)$, we expect power laws for $N(E)$ in 
appropriate regimes of $E$. For example, the very lowest energy behavior is determined by 
the  integral

\be
\int_{E_{min}}^E dE_t\, (E_t-E_{min})^{1/2} (E- E_t)^{-1/2}= (E-E_{min} ) \int_0^1 dy\, [y/(1-y)]^{1/2}
= \frac{\pi}{4}\, (E-E_{min} )
\ee
\label{[9a]}

\be
N(E)= \nu({\bf R}_{min})\, \frac{L\pi}{\hbar}\, (\frac{m}{k_{anis}^3})^{1/2}\, (E-E_{min} )	
\label{9b}
\ee

Note that the prefactor of the energy difference (which we call $a$) is proportional to the volume of 
the nanotubes, so that $N(E)/a$ is an intensive variable. The linear dependence on energy 
above threshold ($E-E_{min}$) is the behavior characteristic of a 4D gas. This initially unexpected result is a logical 
consequence of convoluting a transverse spectrum characteristic of a 3D system  (eq. \ref{8}) with the 
1D longitudinal degree of freedom. Similarly, the behavior at higher energy is 
obtained by convoluting an inverse square  root dependence of $g_{high}(E)$ with the 1D inverse 
square root dependence. Then the high energy result is that $N(E)$  is independent of energy, behavior characteristic of a 2D density of states.
 These analytic results for the spectra are consistent with the numerical results 
seen in Figs. 2 and 3; that is, the low energy density of states is proportional to $(E-E_{min} )$ 
while the high energy behavior is independent of energy. By the word "``high"", we mean 
energies reasonably close to $E_{min}$ (say $E-E_{min}\approx $20mK) 
 but not the absolutely lowest energies.

To summarize this section, we note that unexpected behavior of $N(E)$ emerges from a 
simple model. The low energy spectrum of the nonuniform system is {\it qualitatively 
}
different from the spectrum obtained when heterogeneity is ignored. The latter 
corresponds to 1D physics, i.e. $N(E)$ proportional to $(E - E_{min})^{-1/2}$. With heterogeneity, 
instead, we find that $N(E)$ exhibits a 4D form at very low energy and a 2D form at higher 
energy. 

\section{Prediction of BEC}

In this section we explore the consequences of the anomalous densities of states for 
thermal properties of these systems. In so doing, we assume that the adsorbed gas can 
equilibrate. This entails the rearrangement of the particles as the temperature T is lowered 
by moving from ICs that have high transverse energies to those with lower energies. 
Thus, we assume a combination of sufficient particle mobility and sufficiently patient 
experimentalists that the assumption is valid. In the extreme opposite limit of negligible 
transport of particles the thermal behavior is that of isolated and independent 
1D systems. The difference 
between these behaviors is dramatic.

At very low density, one can ignore the effects of both interparticle interactions and 
quantum statistics. Then, the specific heat $C(T)=dU/dT$ at fixed number $(N_p)$ of particles 
can be computed from the classical Boltzmann expression for the mean energy $U$:

\be
(\frac{U(T)}{N_p})_{classical} = \frac{\int dE\, E\, N(E)\, \exp[-\beta E]}{\int dE\, N(E)]\, \exp[-\beta E]}
\label{[10]}
\ee

Here $\beta^{-1}=k_B T$. One can insert possible power law forms $N(E)\sim E^{(d/2 - 1)}$
 corresponding to 
a gas in $d$ dimensions and derive the familiar relation

\be
(\frac{C}{ N_p k_B})_{classical} = \frac{d}{2}
\label{[11]}	
\ee

Fig. 4 presents the resulting classical behavior for the systems of He and H$_2$ in nanotubes. One 
observes that the low $T$ regime does correspond to the expected 4D gas behavior, 
$C/( N_p k_B)=2$ while the high T regime exhibits 2D gas behavior, $C/( N_p k_B)=1$. The initial 
rise above 2  at very low $T$  is explained by a small upward deviation from linearity found in $N(E)$ at low $E$. That is, if $N(E)$ at 
low $E$ is proportional to $E(1+E/E_1)$, where $E_1$ is some constant, then $[C/( N_p k_B)]_{classical} = 2 [1+2/(\beta E_1)]$. The 
crossover between the 4D and 2D regimes is gradual, beginning at a few mK and 
complete by 20 mK; the latter value corresponds to the energy above which $N(E)$ is 
virtually identical to the 2D form.

At higher density, one must consider  the effects of interactions and quantum 
statistics. In the present treatment we ignore interparticle interactions, except for some 
comments in the next section. While this is a common assumption, usually adopted for 
simplicity, there exists some justification for it in the present instance because of 
electrodynamic screening of the van der Waals interaction by the surrounding medium; 
there are additional elastic screening effects that have yet to be studied \cite{37}. We do not 
know the resulting effective interaction well enough to characterize it. In contrast, the 
effects of quantum statistics are easily evaluated by the standard procedures. In particular, 
the assumed constraint of fixed particle number yields an implicit relation for the 
chemical potential $\mu$, which depends on $T$, from which one may derive the energy:

\be
N_p = \int dE\, n(E,T)
\ee
\label{[12]}	

\be
U(T) = \int  dE\, E\, n(E,T)
\ee
\label{[13]} 

\be
n(E,T) =N(E)\, (\exp[\beta (E -\mu)]- 1)^{-1}
\ee
(As usual, the classical equations \ref{[10]} and \ref{[11]} are obtained in the limit when the fugacity 
$Z= \exp(\beta\mu) \ll 1$). From $U(T)$, one may compute $C(T)$. In evaluating these formulae, one 
encounters a phenomenon that is well known to occur in 3D bose gases: BEC. This 
transition is derived here in the usual way, by evaluating the implication for the variable $\mu$ (at 
fixed $N_p$) as $T$ is lowered: upon decreasing $T$, $\mu$ increases up to the point where it reaches 
its highest possible value $\mu=E_{min}$. Thus, the BEC condition for the transition 
temperature $T_c=(k_B \beta_c)^{ -1}$ is

\be
N_p = \int dE\, N(E) \, (\exp[\beta_c (E - E_{min})] - 1)^{-1}
\ee
%
Since $\mu$  cannot exceed $E_{min}$, as $T$ falls below $T_c$
 (given implicitly by this relation), there 
ensues a division into two groups of particles- the $N_{excited}$ particles accounted for by the 
integral over excited states $(E> E_{min})$ and the remaining particles having $E= E_{min}$ that 
comprise the bose condensate, $N_{cond}$ : 

\be
N_{cond} = N_p - N_{excited} \;\;		(T< T_c)		
\ee

\be
N_{excited} = \int dE\,  N(E) (\exp[\beta (E - E_{min})] - 1)^{-1}
\ee
The results of these calculations appear in Figs. 5 to 8.
Fig. 5 depicts the dependence of $Z$ on the relative temperature $T/T_c$ for two different 
densities. Fig. 6 displays the dependence on $T$ of the condensate fraction $f(T) = N_{cond} / 
N_p$. Fig. 7 shows the function $C(T)$, which exhibits singular behavior at the transition. 
Finally, Fig. 8 shows the dependence of $T_c$ on $N_p$. Note that the transition temperature 
falls within the mK range studied in  many low temperature laboratories, so experimental 
observation of this transition is feasible.

In the preceding three figures, comparison is made with results obtained for a purely 4D 
bose gas defined to have the same initial slope of the density of states as the real system. 
That ideal system satisfies $N(E)= a E$, for all $E>0$, with the analysis presented in the 
Appendix. (For the real system, the $a$ coefficient is the prefactor of the energy term in Eq. 
\ref{9b}). The key results for the 4D system are

\be
f_{4D}(T) =1 - (\frac{T}{ T_{c,4D}})^2
\ee

\be
C_{4D}(T)= \frac{36 \zeta(3)}{\pi^2} (\frac{T}{ T_{c,4D}})^2,  \;  T<T_{c,4D}
\label{18}
\ee

\be
T_{c,4D} = (\frac{6N_p}{a})^{1/2}\frac{1}{\pi k_B}
\label{19}
\ee
Here $\zeta(3)=1.202$ is a Riemann zeta function. The $T^2$ dependences of $f(T)$ and $C(T)$ are 
exhibited by the ``real" system at low $T$, as seen in the figures. At the very lowest 
temperatures, only the lowest energy states play a role; since $N(E)$ is the same for the ideal system 
in this 
region, the behaviors coincide. The transition temperature is higher for the real system 
than for the ideal one because at high $E$, $N(E)$ is lower in the real case than in the ideal 
case, implying that (for the same value of $N_p$) $\mu$ reaches its transition value $E_{min}$ at a 
higher value of $T_c$.

Since the ground state energy of the noninteracting classical system coincides with that of 
the noninteracting quantum system, as does the energy at very high $T$, it follows that the 
integral of the two heat capacities is the same. 
 At any particular $T$, of course, the values of $C$ are very different. In particular, the 
quantum system manifests the reduced $C$ at low $T$ expected in this bose gas case. The 
quantum system exhibits a much larger value of C in the vicinity of $T_c$, which is 
dominated by a large rate of excitation out of the condensate; see Fig. 7.

One interesting question is this: how are the particles distributed in energy as a function 
of $T$. To address this, we present in Fig. 9 the occupied energy distribution function
$n(E,T)$ at various temperatures. One observes the increasing relative occupation of low 
energy states with decreasing $T$; for $T>T_c$, their sum is just $N_p$, the total number present. 
At $T_c$ there occurs a discontinuous jump in the occupation of the lowest energy states 
accompanying the BEC transition. Below $T_c$, the number $N_{excited}$ of uncondensed particles 
decreases, vanishing at $T=0$, when all of the particles settle into the condensate.

Our prediction of BEC in this anisotropic system raises a question: what is the relationship between the transition condition and that of more familiar homogeneus problems, e.g. the 3D Bose gas and the Kosterlitz-Thouless transition of a superfluid film? In the latter examples, the relevant criterion is  if the form $\rho\lambda^d\sim 1$, where $\rho$ is the number density, $d$ the dimensionality and $\lambda=(2\pi\beta\hbar^2/m)^{1/2}$ is the de Broglie thermal wavelength. We present now a heuristic argument  that is consistent with this homogeneus relation and yields a good estimate of the transition temperature for the present BEC problem. The argument estimates $T_c$ by analyzing the condition for quantum degeneracy; the number of particles $N_p$ becomes comparable to the number of states $N_{states}(T_c)$ lying  within $k_BT_c$ of the ground state. The number of states within that interval is essentially the product of the number of states of longitudinal motion within that range and the number of states of transverse motion within that range:

\begin{eqnarray}
N_{states}(T)&\sim& N_{\|}(T)\;N_{\bot}(T)\nonumber\\
&\sim& (\frac{(mk_BT)^{1/2}}{\hbar/L}) (\frac{\nu}{(\beta k_{anis})^{3/2}})\nonumber
\end{eqnarray}
The factor $N_{\|}$ is the ratio of the thermal momentum in the $z$ direction  to the spacing between discrete states. The factor $N_{\bot}$ is derived from a thermal spread of states in ${\bf R}$ space [$\delta R\sim (\beta k_{anis})^{-1/2}$] and the corresponding number of states in the experimental sample, $N_{\bot}\sim \nu \delta R^3$. Setting $N_p = N_{states}(T_c)$ yields our estimate of the BEC transition condition:

\be
N_p \sim \frac{L\nu}{\hbar\beta^2}\frac{m^{1/2}}{k_{anis}^{3/2}}
\ee
Apart from numerical factors, this result coincides with the combination of Eqs. 10 and 21. 
Note that this last equation can be rewritten in the  form 
$   \rho^* \lambda^4 \sim 1$ with  $\rho^*= N_p/(L \;l^3)$  where $l= \nu^{1/3}\hbar/ (m \;k_{anis})^{1/2}$ is a characteristic length defined by the energetic  parameters and the distribution of radii.

\section{Summary and discussion}

In this paper we describe a conceptually simple system that turns out to exhibit 
remarkable behavior. The focus of the study is the lowest-lying states of a system of 
hydrogen or helium moving within a collection of nanotubes having many ICs. Our 
analysis yields a density of states $N(E)$ that fulfills the requirement of BEC (i.e. the 
maximum integrated occupation number is finite). Equally interesting, perhaps, is the 
finding that the calculated properties at low $T$ are characteristic of a 4D gas in free space. 
Since (to the best of our knowledge) no 4D gas has been observed previously, the 
predicted low T and critical behaviors of this system are particularly interesting.

A number of theoretical issues need to be discussed. First, one must ask whether the ideal 
gas assumption is valid; indeed, our group has previously explored the vapor-liquid 
transition of these gases using quasi-1D models of perfect ICs \cite{38}. A concern is 
therefore that such condensation in a somewhat disordered system would either preempt 
the BEC or alter the nature of the BEC, as is the situation with 3D superfluid $^4$He. 
However, it may well be that disorder reduces the temperature of the hypothetical condensation 
transition below that of the BEC transition, so that two separate transitions occur. 

A related question is the role of screening by the medium of the interparticle interaction. 
On the bare surface of graphite, experimental and theoretical evidence has indicated that 
the well-depth is reduced by some 10 to 20\% \cite{39}. Intuition and some calculations 
\cite{37} suggest that the attraction would be reduced by a much greater factor for gases 
within a nanotubes array. That is consistent with the neglect of these interactions in the 
present paper.

The requirements for observation of this phenomenon in the laboratory are not trivial to 
satisfy. A particularly serious concern is that equilibrium is difficult to achieve. It 
requires particles to diffuse out of one IC and into another as $T$ changes. The slow 
equilibration is a potentially fatal problem that is difficult to assess; experimental 
diffusion data would be very helpful in this regard.\cite{Narehood,Skoulidas}
 We note that one need not imagine 
that particles are required to move macroscopic distances; one can employ tubes of quite 
finite length in order to accelerate the process of equilibration. Evidently, one should not 
underestimate this problem, which is an important concern in much broader contexts involving 
nanotubes, such as gas storage and isotope separation. 

The problem at hand raises the venerable question of the relation between BEC and 
superfluidity (SF). \cite{footnote2} We know that BEC and SF appear simultaneously at the lambda 
transition of $^4$He \cite{3,4}. However, the Kosterlitz-Thouless transition of 2D films is one 
involving SF without BEC \cite{40}. Moreover, the 3D ideal bose gas exhibits BEC without 
SF, since the Landau velocity criterion is not satisfied. The latter says that the threshold 
velocity for superflow is the minimum of the ratio of the excitation energy to the 
momentum \cite{41}. Our unusual nanotubes-hosted BEC state would also fail this test, 
implying that no superflow is possible. Another aspect of SF is quantized circulation, 
which is impossible in this environment. Thus, we think that  SF is unlikely to 
accompany the  transition described above. If, however, one were to include weak interparticle interactions, SF could occur.

The ideal nanotube radii for the lowest energy states fall in the range 0.8 to 1 nm, which 
is an advantageous size because typical samples  contain many such tubes. If, for 
example, only smaller tubes are present in a particular sample, one would have to explore 
the behavior without all of the simplifying mathematics employed here. We have yet to 
tackle that problem in general, but do note some interesting results found in one case. 
Suppose for simplicity that the transverse density of states for a given sample (consisting 
of N$_{IC}$ ICs) has a form (that is quantitatively wrong but qualitatively reasonable in that it 
consists of a spread of transverse energies) given by the ""uniform barrier"" function:

\be
g(E_t)=\frac{N_{IC}}{(E_> - E_<)}	,   E_< < E_t < E_> \nonumber
\ee
\be
g(E_t)=0		,  otherwise \nonumber
\ee
The result in this case (from Eq. \ref{7}) is 

\be
N(E) = \frac{2 N_{IC}}{(E_>- E_<)}\,  \frac{L}{\pi\hbar}\, (\frac{m}{2})^{1/2}  ((E-E_<)^{1/2} \Theta[E- E_<] - (E-E_>)^{1/2}
 \Theta[E- E_>] )
\ee
This expression yields the 1D limiting case of Eq. 4 when the barrier width $(E_>- E_<)$ falls 
to zero. In the case of nonzero width, this expression yields 3D behavior for $ N(E)$  at low 
E and 1D behavior at high $E$. The reason for the 3D limit is that the constant transverse 
density of states in the finite uniform barrier model mimics the constant density of states 
of a 2D gas. In this model, one derives a BEC transition from this form of $ N(E)$, with 
thermal behavior quite different from that described elsewhere in this paper, i.e., when 
the lowest possible transverse energies (near $ E_{min}$) play the dominant role in determining 
the behavior. The transition temperature of a system of $N_p$ molecules, obtained for the 
uniform barrier model with the conventional 3D theory, satisfies 

\be
( k_B T_c)^3 = \frac{2 \pi}{m} (\frac{N_p (E_>- E_<) \hbar}{L N_{IC}\zeta(3/2)})^2
\ee

This proportionality of $T_c$ to $N_p^{2/3}$  is characteristic of 3D behavior, in contrast with the 4D 
proportionality to $N_p^{1/2}$ in Eq. \ref{19}. 

One might ask whether similar BEC phenomena occur in a related environment, the 
groove between two nanotubes (e.g. at the external surface of a bundle). Adsorption in this region has been much studied in 
theory and experiments  \cite{42}. We have therefore investigated the consequence 
of heterogeneity for that problem, with interesting results. One finding is that the 
symmetric situation (two identical radii) provides a local {\it maximum} (in 2D {\bf R} space) of the 
transverse energy. Hence, the methodology and results of the present paper are not 
applicable, for the most part. The one related result is that there occurs in the groove case 
an anomalous signature of inhomogeneous broadening in the excitation spectrum. 
This is qualitatively analogous to the behavior of $N(E)$ here, except that spectroscopy is 
the tool of choice for its investigation.\cite{17}

In closing, we emphasize an intriguing fact: this transition is a consequence of the 
nonuniformity of the ICs within experimental samples. A perfectly uniform set of ICs 
would exhibit 1D physical properties instead of the transition described here. Such a 
remarkable consequence of heterogeneity has precedents in other physical systems, of 
which the spin glass phenomenon is a well-known example \cite{43,44,45}. 

We are grateful to Carlo Carraro, Vincent Crespi, Allan Griffin, Susana Hern\'andez, Paul Lammert, Ari Mizel, 
Aldo Migone, Paul Sokol and Flavio Toigo for helpful discussions. This research has 
been supported by the National Science Foundation. F.A. acknowledges funding from MIUR-COFIN 2001.

\appendix*
\section{4D ideal Bose gas}

We 
consider  an ideal  bose gas in four dimensions whose density of states $N(E)$ satisfies

\be
N(E)=  a\; E  \;\;(E>0)
\ee

For this system, we evaluate the number of particles,  energy and specific heat 
 in the usual way, using Eqs. 13-15.
 Straightforward calculations yield 

\be
N_p=a \;g_2(Z)\; (k_B T)^2 + N_{cond}
\label{nbec}
\ee
where $N_{cond}$ is the number of bose  condensed particles, and  
\be
U= 2\; a \;g_3(Z)\;(k_BT)^3. 
\ee
Here the functions $g_2$ and $g_3$  are second and third "Bose Einstein integrals",

\be
g_n(Z) = \frac{1}{\Gamma(n)}\int_0^{\infty}    \frac{x^{n-1}dx}{Z^{-1}e^x-1}.
\ee

The fugacity $Z(T)$ is determine from Eq. \ref{nbec}. Moreover,  from this equation we obtain the critical temperature $T_c$ for a given number of particles $N_p$,  given that the maximum value of $Z$ is 1 and $g_2(Z)$ is a monotonically  increasing function, 

\be
k_BT_c=\sqrt{\frac{N_p }{\zeta (2) a}}
\label{tc}
\ee
From Eqs. \ref{tc} and \ref {nbec} the condensate fraction results,
\be
f=\frac{N_{cond}}{N_p} = 1-(\frac{T}{T_c})^2.
\label{f}
\ee
Now  we evaluate the specific heat. To do so we have to be aware that the fugacity depends on the temperature only for $T>T_c$ while it is equal to 1 for $T\le T_c$. Then we obtain

\be
\frac{C_N}{N_pk_B} =   6\;\frac{g_3(Z)}{g_2(Z)}-4\;\frac{g_2(Z)}{g_1(Z)} \,\,(T\ge T_c)\nonumber
\ee
\be
 =   6\; \frac{\zeta(3)}{\zeta(2)} (\frac{T}{T_c})^2 (T \le T_c)\nonumber
\ee
Note that $C_N$ is a continuous function of T, with  
 its value at the critical temperature $C_v(T_c) =  4.4 N_p k_B$.

In the these derivations we have used some properties of the functions $g_n(Z)$,
\be
g_n(1)=\zeta(n)
\ee
and
\be
Z g'_n(Z) = g_{n-1}(Z).
\ee
These, together with equation A.2, give the following relationship,

\be
\frac{\partial g_2(Z)}{\partial T} = -\frac{2}{T} \;g_2(Z)
\ee

\newpage

\begin{table}[tbh]
\begin{tabular}{|c|c|c|c|c|} \hline

	&$\sigma$ (\AA) &$\epsilon$ (K) &  $R_{min}$ (\AA) & $E_{min}$ (K)\\
\hline\hline
H$_2$ $^1$ & 2.97& 42.8 & 8.75 & -1159.2\\
H$_2$ $^2$ & 3.23 & 32.19 & 9.95 & -1052.4 \\
$^4$He $^3$ &  2.74 & 16.2 & 7.19 & -353.6 \\
$^4$He $^2$ & 2.98 & 16.9 & 8.48 & -456.9 \\
\hline

\end{tabular}

\caption{ Numerical values of $E_{min}$ and $R_{min}$, defined in the text. 
The parameters of the potential used in each case are from Refs. $^1$  \cite{gordillo},  $^2$ \cite{stanprb} and $^3$ \cite{stan}}

\end{table}

\newpage

FIGURE CAPTIONS

1. Contours of the ground state transverse energy $E_t({\bf R})-E_{min}$ (relative to its minimum 
value) of an H$_2$ molecule as a function of the radii ${\bf R}=(R_1,R_2,R_3)$ of the tubes bordering 
the IC. $ {\bf R}$ is measured relative to the minimum ${\bf R}_{min}$ ; the figure depicts behavior along 
the line $(0,1,-1)$. Beginning with the contour closest to the minimum (the closed curve), 
the contours correspond to relative energy 0.005, 0.01, 0.05, 0.1, 0.5, 1 and 5 K.

2. Upper panel: transverse density of states  $g(E)$ for $^4$He, computed from the distribution 
of ICs as described in the text. Lower panel: total density of states $N(E)$. In both cases, 
the energy is measured relative to the minimum $E_{min} =-456.88K$

3. Transverse and total densities of states, as in Fig.2, for H$_2$. $E_{min} = -1052.97K$

4. Specific heat of He (full curve) and H$_2$ (dots) as a function of $T$ computed using the 
classical Boltzmann approach.

5. Fugacity $Z=\exp(\beta\mu)$ as a function of $T/T_c$ for two different densities: 1 x  10$^{-5}$ \AA$^{-3}$ (full 
curve) and 0.5 x 10$^{-5}$ \AA$^{-3}$ (dots) within an inhomogeneous nanotube array while the 
dashed curve corresponds to an ideal 4D gas having the same low energy density of 
states.

6. Condensate fraction as a function of the temperature computed for  H$_2$ in the nanotube system (full curve) and the 4D H$_2$ gas 
(dots) for $N$=1x10$^{-5}$ \AA$^{-3}$.

7. Heat capacity for  H$_2$ as a function of the temperature in 
nanotubes (full curve) and for the  4D H$_2$ gas (dashed) (panel a) and for  $^4$He in 
nanotubes (full curve) and for the  4D $^4$He gas (dashed) (panel b).

8. Density of H$_2$ molecules (full curve) and $^4$He atoms (dotted curve) as a function of 
transition temperature $T_c$. The short-dash (H$_2$) and long dash (He) curves indicate the 4D 
ideal gas
limiting behavior.

9. The occupied state energy distribution $n(E,T)$ (relative to the total number $N_p$)
 defined in Eq. 15 is shown for H$_2$ at relative temperatures $T/ T_c 
=0.5$ (full curve),1 (dots), 1.1 (short dashes) and 1.5( long dashes). The integral of the 
curve at 0.5 is the fraction (about 0.7) of excited particles while the other curves integrate 
to 1, since then the condensate fraction vanishes.

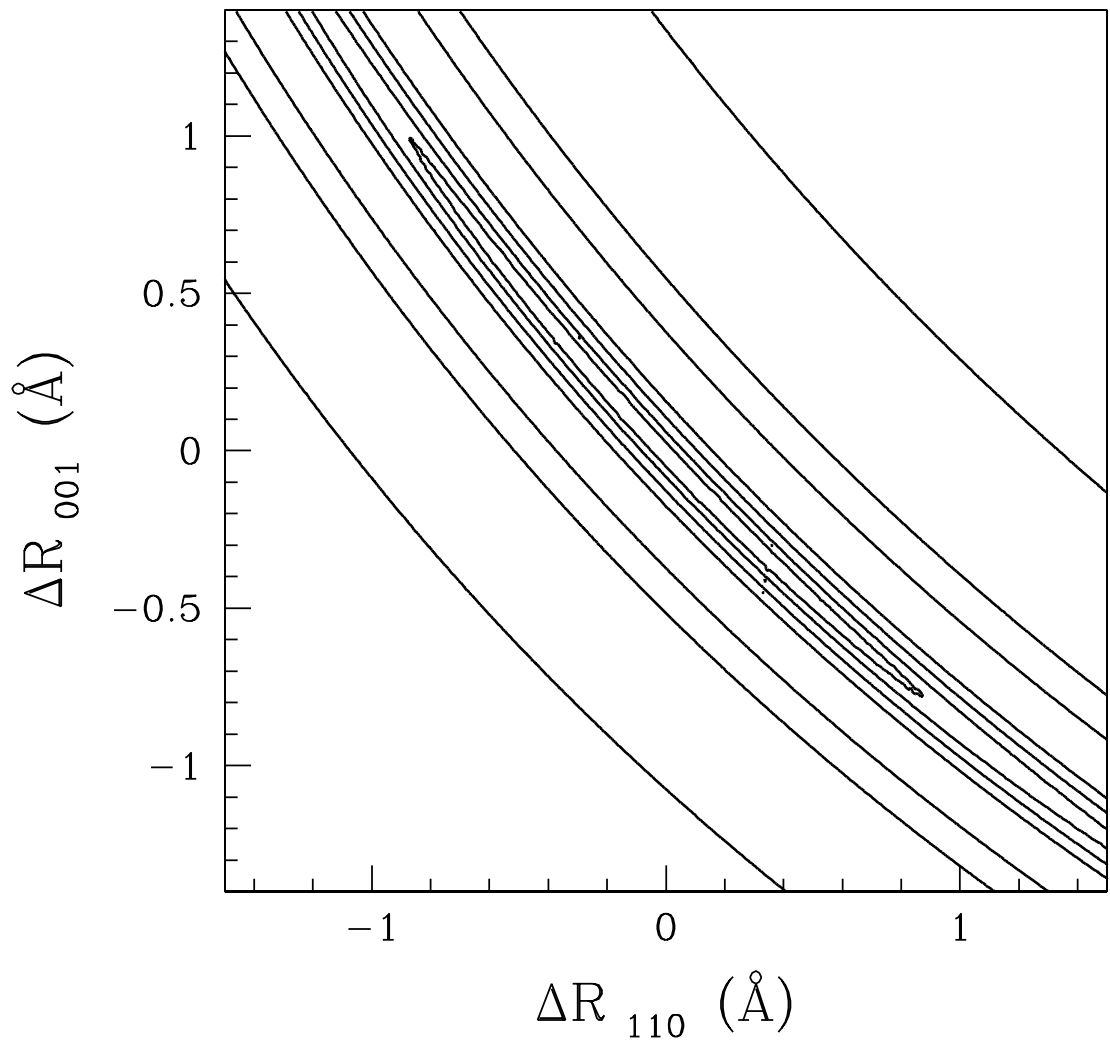
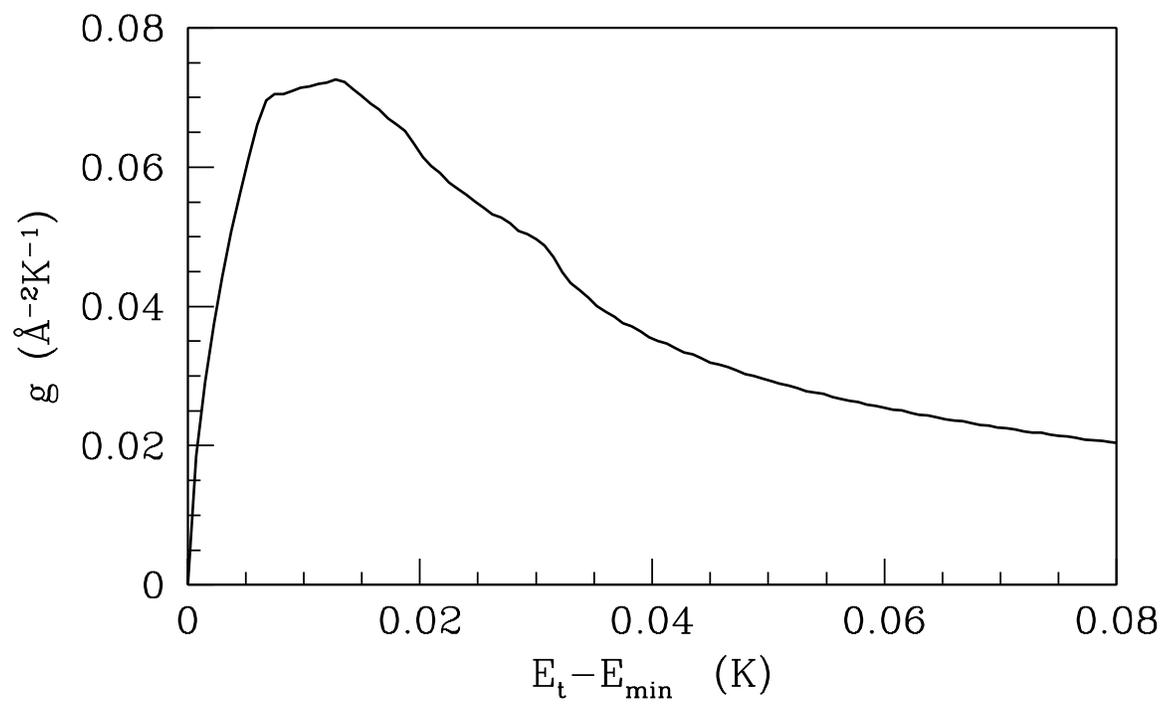
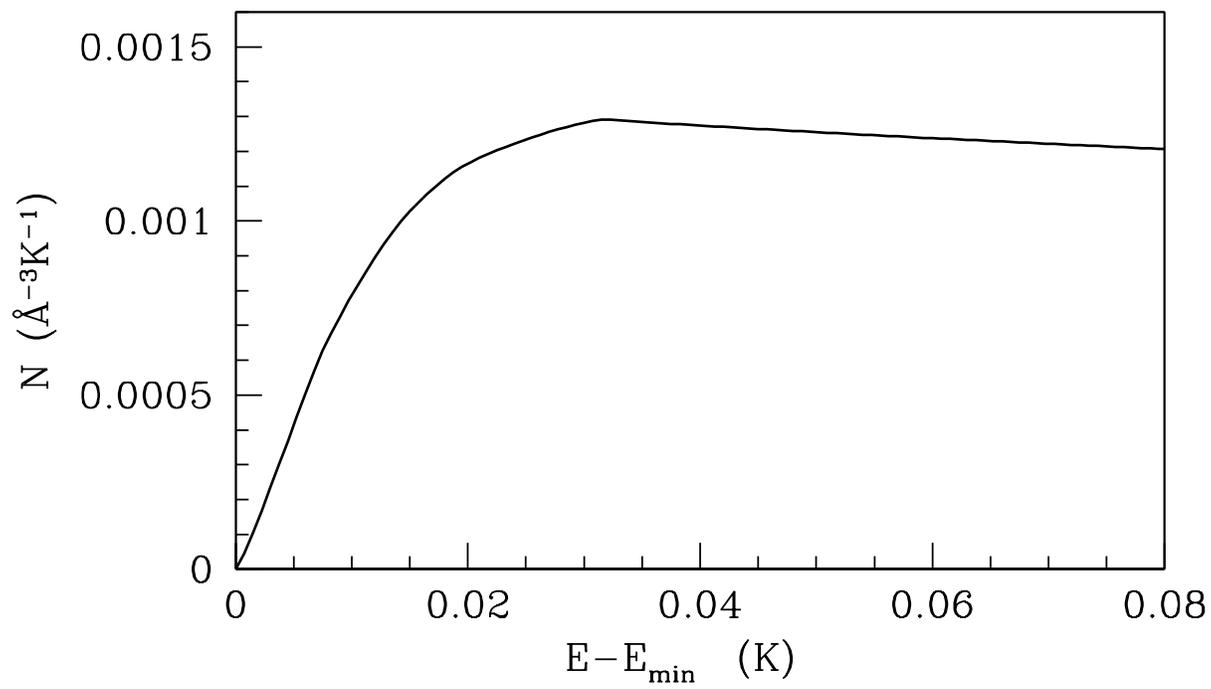
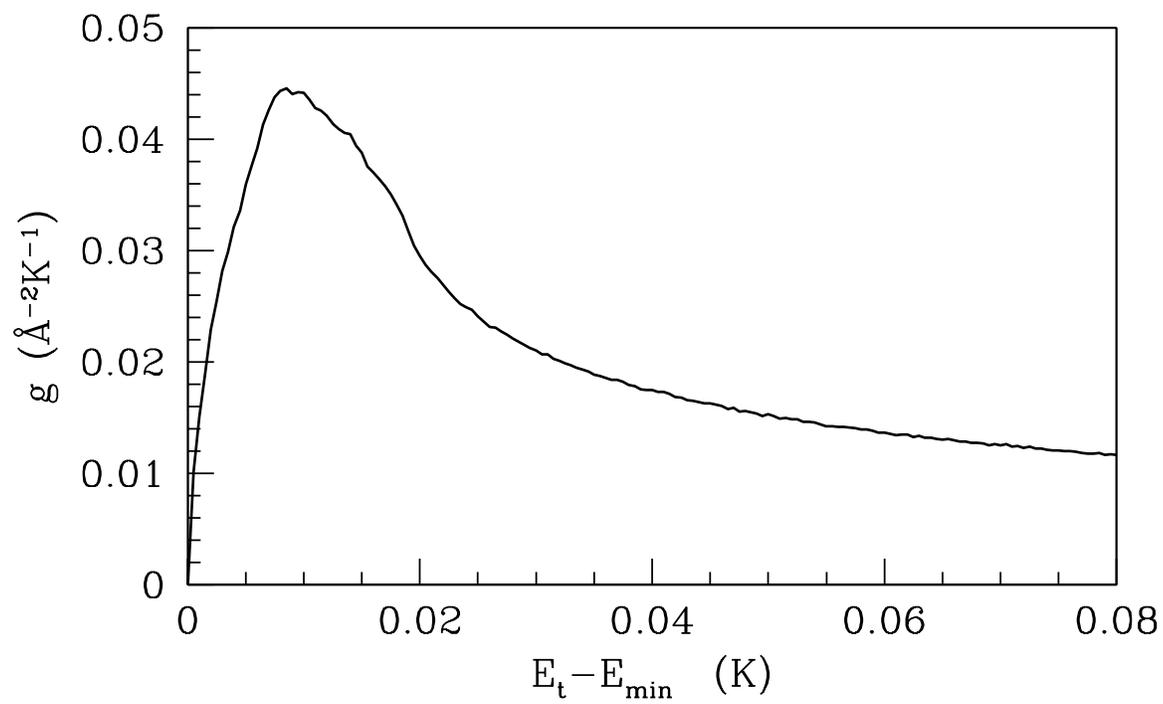
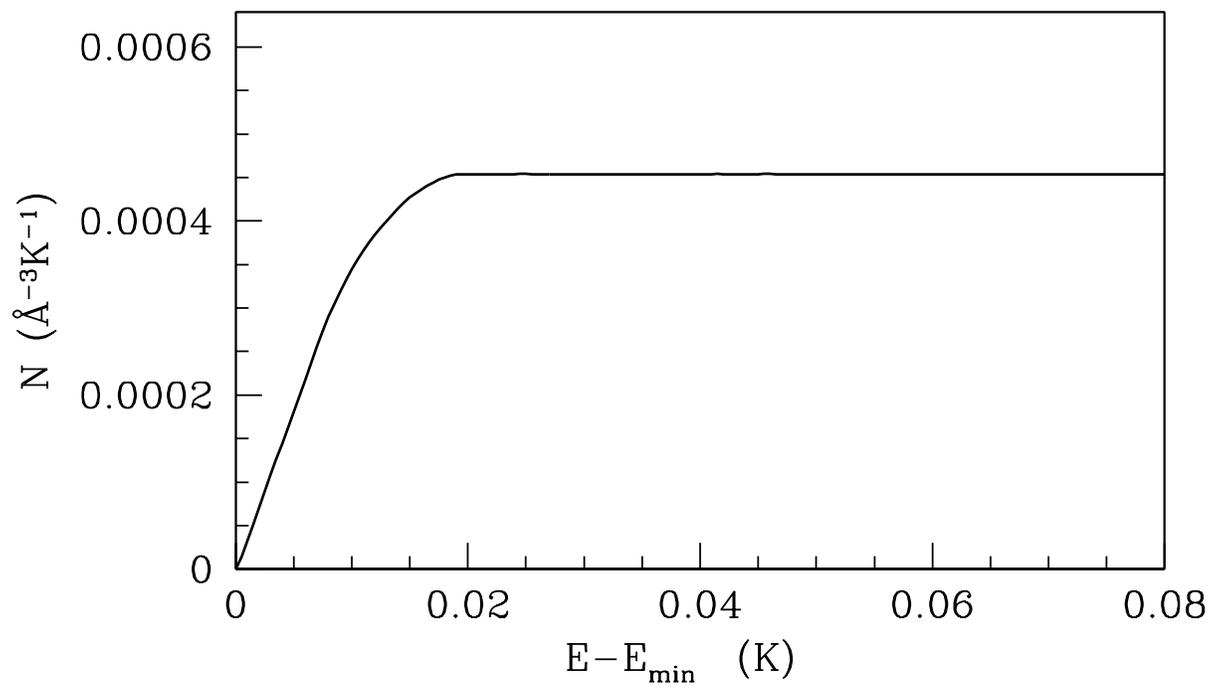
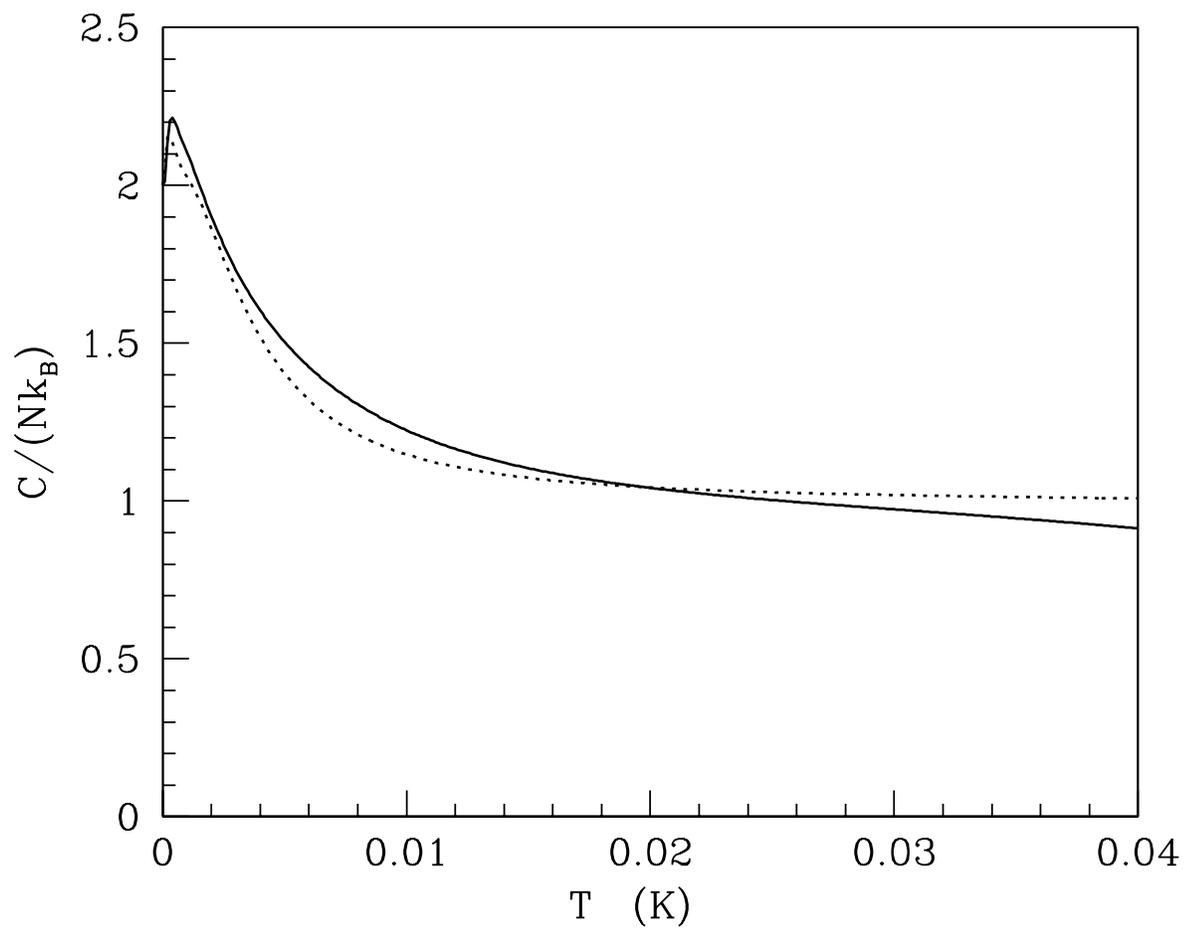
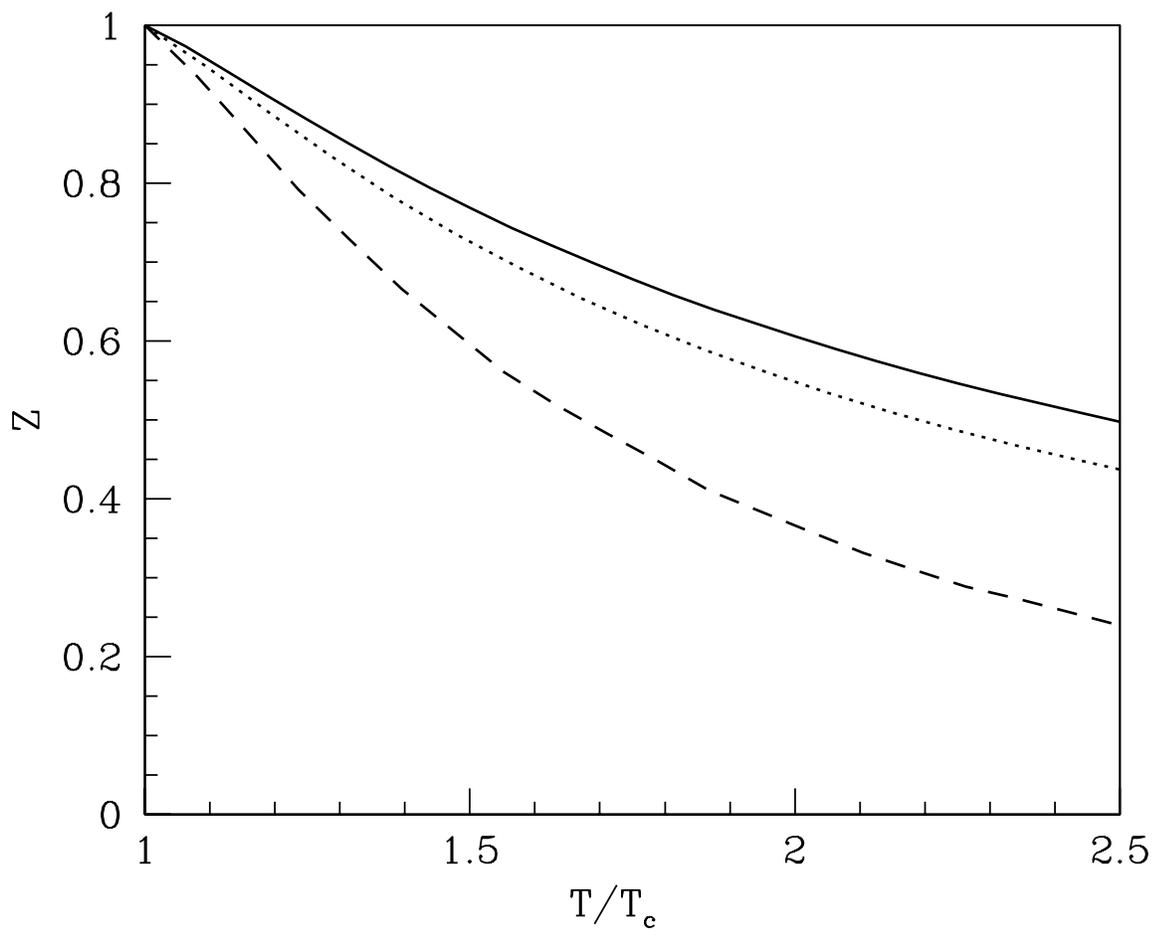
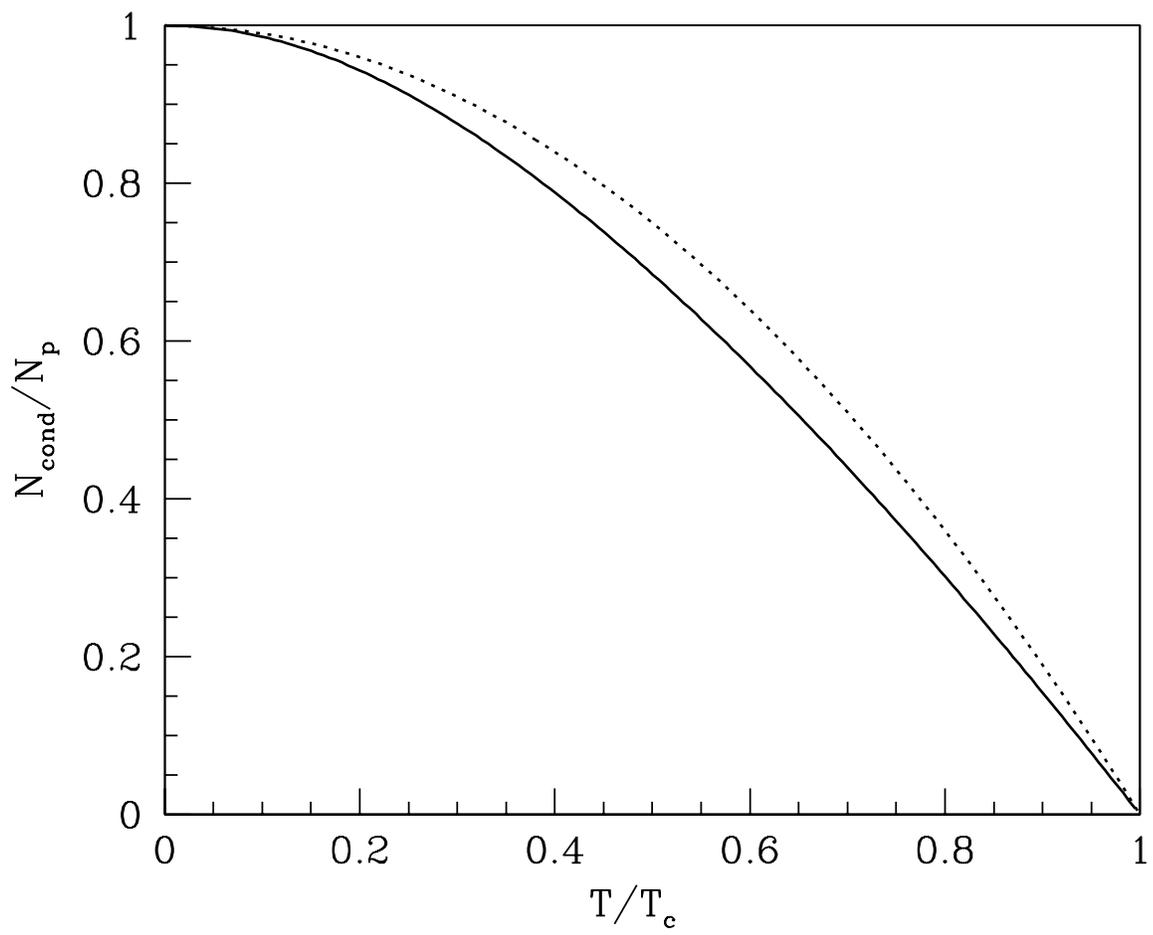
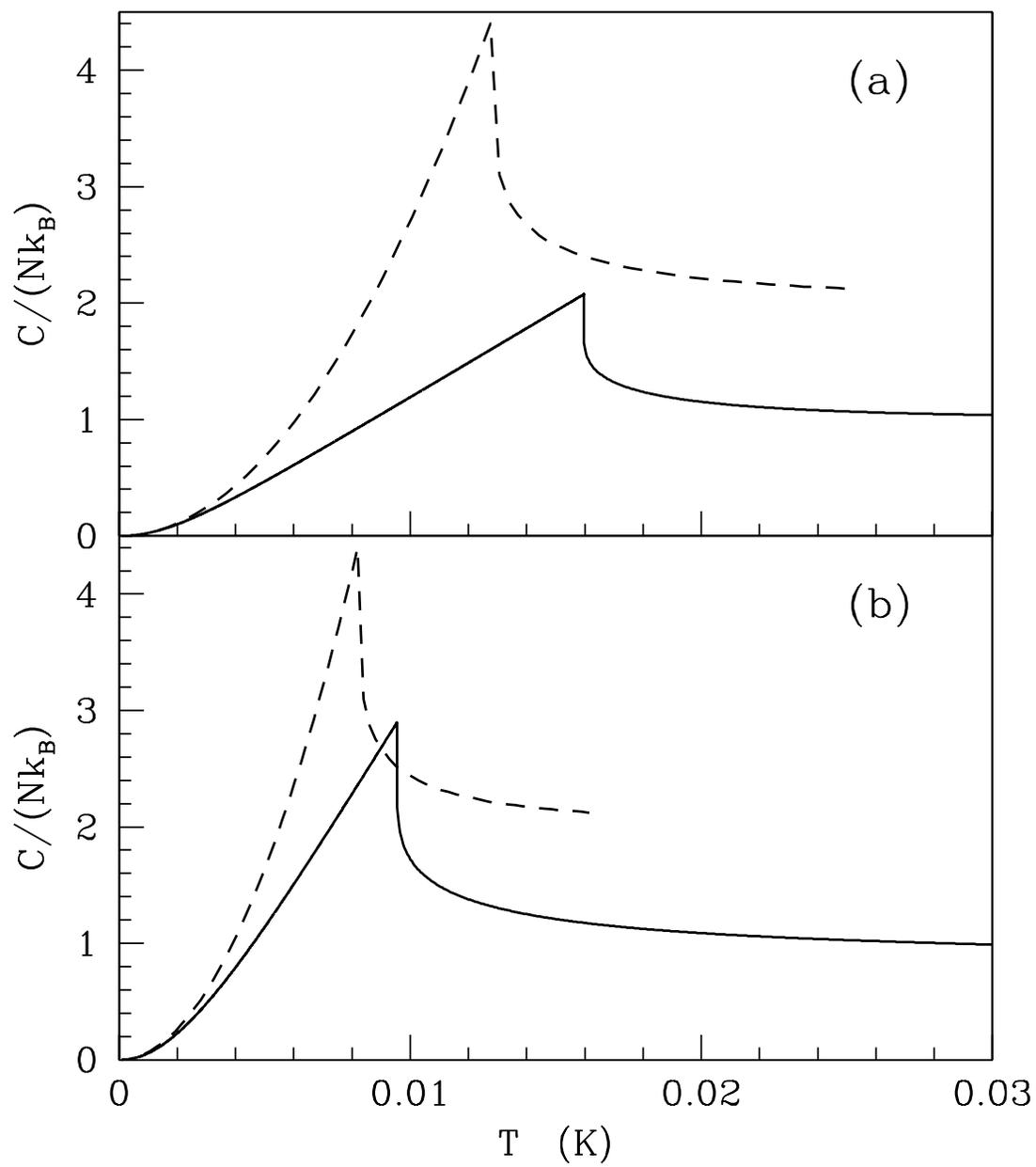
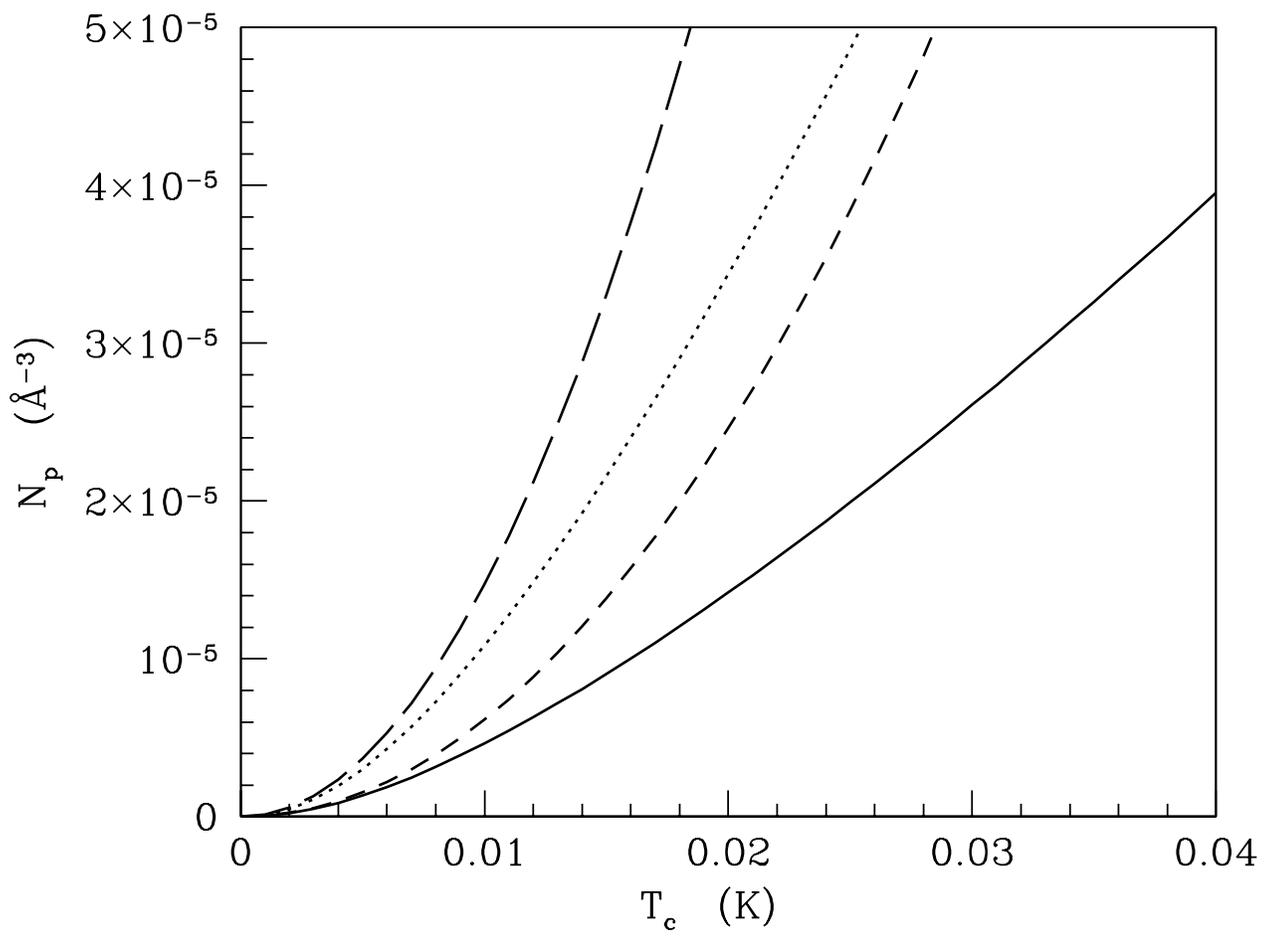
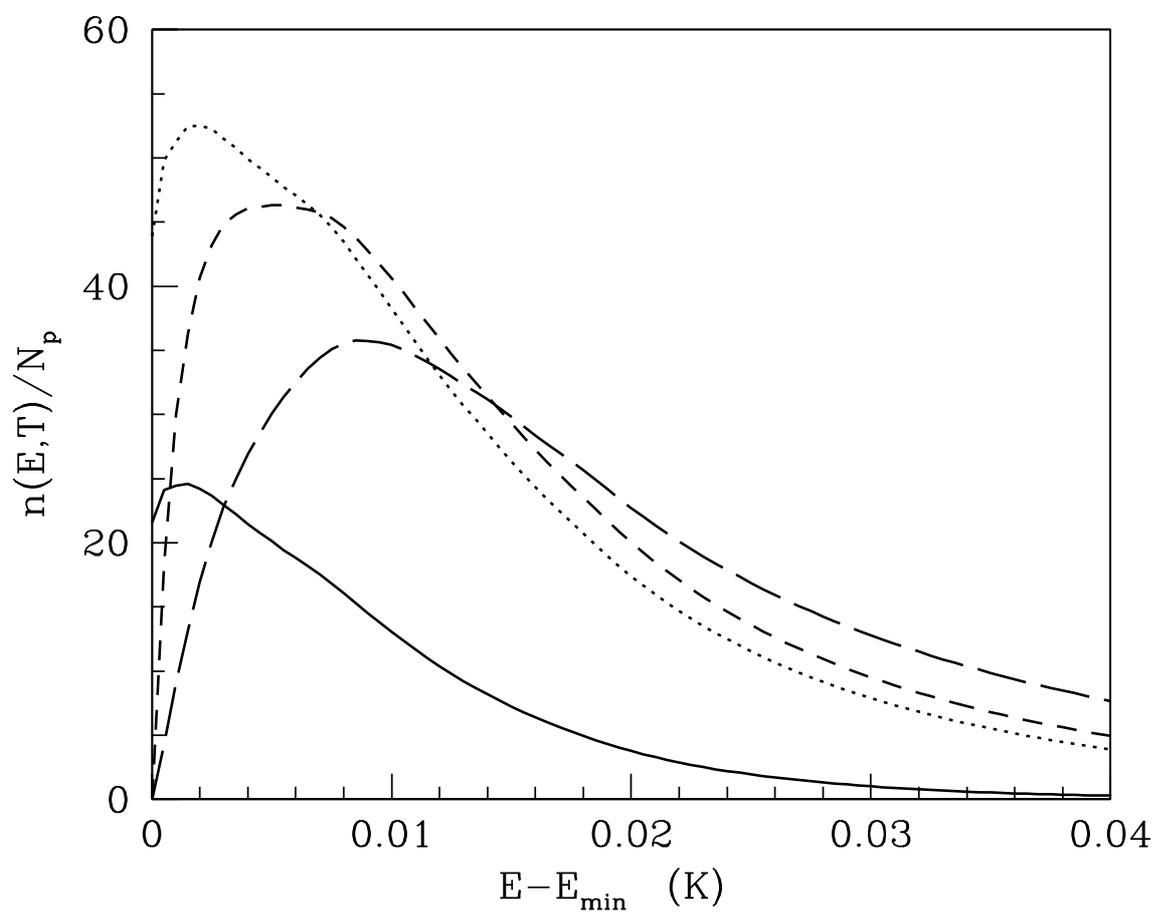
\end{document}